\documentclass[prd,nofootinbib,preprint,superscriptaddress]{revtex4}

\usepackage[T1]{fontenc}
\usepackage[latin9]{inputenc}
\usepackage{amsmath}
\usepackage{amsthm}
\usepackage{amssymb}
\usepackage{cancel}
\usepackage{graphicx}
\usepackage[unicode=true, bookmarks=false, breaklinks=false,pdfborder={0 0 1},backref=section,colorlinks=false] {hyperref}
\usepackage{fancyhdr}
\usepackage{epsfig}
\usepackage{slashed}
\usepackage{mathrsfs}
\usepackage[thinc]{esdiff}
\usepackage[titletoc]{appendix}
\usepackage{tikzsymbols}
\usepackage{float}
\usepackage{feynmf}
\usepackage{tikzsymbols}
\usepackage{dsfont}
\usepackage{xcolor}
\usepackage{slashed}
\usepackage{braket}
\usepackage{mathrsfs}
\usepackage{xcolor}
\usepackage{gensymb}
\usepackage{latexsym}
\usepackage[compat=1.1.0]{tikz-feynman}
\newcommand{\arXivold}[2]{\href{http://arxiv.org/pdf/#1}{{\tt #2/#1}}}

\definecolor{lime}{HTML}{A6CE39}
\DeclareRobustCommand{\orcidicon}{
	\begin{tikzpicture}
	\draw[lime, fill=lime] (0,0)
	circle [radius=0.2]
	node[white] {{\fontfamily{qag}\selectfont \tiny ID}};
	\draw[white, fill=white] (-0.0625,0.095)
	circle [radius=0.007];
	\end{tikzpicture}
	\hspace{-2mm} }
\foreach \x in {A, ..., Z}{\expandafter\xdef\csname orcid\x\endcsname{\noexpand\href{https://orcid.org/\csname orcidauthor\x\endcsname}
			{\noexpand\orcidicon}} }

\definecolor{lime}{HTML}{A6CE39}

\foreach \x in {A, ..., Z}{\expandafter\xdef\csname orcid\x\endcsname{\noexpand\href{https://orcid.org/\csname orcidauthor\x\endcsname}
			{\noexpand\orcidicon}} }

\makeatother
\RequirePackage[normalem]{ulem} 
\RequirePackage{color}\definecolor{RED}{rgb}{1,0,0}\definecolor{BLUE}{rgb}{0,0,1} 
\providecommand{\DIFadd}[1]{{\protect\color{blue}\uwave{#1}}} 
\providecommand{\DIFaddbegin}{} 
\providecommand{\DIFaddend}{} 
\providecommand{\DIFdelbegin}{} 
\providecommand{\DIFdelend}{} 
\providecommand{\DIFaddbeginFL}{} 
\providecommand{\DIFaddendFL}{} 
\providecommand{\DIFdelbeginFL}{} 
\providecommand{\DIFdelendFL}{} 
\newcommand{\DIFscaledelfig}{0.5}
\RequirePackage{settobox} 
\RequirePackage{letltxmacro} 
\newsavebox{\DIFdelgraphicsbox} 
\newlength{\DIFdelgraphicswidth} 
\newlength{\DIFdelgraphicsheight} 
\LetLtxMacro{\DIFOincludegraphics}{\includegraphics} 
\newcommand{\DIFaddincludegraphics}[2][]{{\color{blue}\fbox{\DIFOincludegraphics[#1]{#2}}}} 
\newcommand{\DIFdelincludegraphics}[2][]{
\sbox{\DIFdelgraphicsbox}{\DIFOincludegraphics[#1]{#2}}
\settoboxwidth{\DIFdelgraphicswidth}{\DIFdelgraphicsbox} 
\settoboxtotalheight{\DIFdelgraphicsheight}{\DIFdelgraphicsbox} 
\scalebox{\DIFscaledelfig}{
\parbox[b]{\DIFdelgraphicswidth}{\usebox{\DIFdelgraphicsbox}\\[-\baselineskip] \rule{\DIFdelgraphicswidth}{0em}}\llap{\resizebox{\DIFdelgraphicswidth}{\DIFdelgraphicsheight}{
\setlength{\unitlength}{\DIFdelgraphicswidth}
\begin{picture}(1,1)
\thicklines\linethickness{2pt} 
{\color[rgb]{1,0,0}\put(0,0){\framebox(1,1){}}}
{\color[rgb]{1,0,0}\put(0,0){\line( 1,1){1}}}
{\color[rgb]{1,0,0}\put(0,1){\line(1,-1){1}}}
\end{picture}
}\hspace*{3pt}}} 
} 
\LetLtxMacro{\DIFOaddbegin}{\DIFaddbegin} 
\LetLtxMacro{\DIFOaddend}{\DIFaddend} 
\LetLtxMacro{\DIFOdelbegin}{\DIFdelbegin} 
\LetLtxMacro{\DIFOdelend}{\DIFdelend} 
\DeclareRobustCommand{\DIFaddbegin}{\DIFOaddbegin \let\includegraphics\DIFaddincludegraphics} 
\DeclareRobustCommand{\DIFaddend}{\DIFOaddend \let\includegraphics\DIFOincludegraphics} 
\DeclareRobustCommand{\DIFdelbegin}{\DIFOdelbegin \let\includegraphics\DIFdelincludegraphics} 
\DeclareRobustCommand{\DIFdelend}{\DIFOaddend \let\includegraphics\DIFOincludegraphics} 
\LetLtxMacro{\DIFOaddbeginFL}{\DIFaddbeginFL} 
\LetLtxMacro{\DIFOaddendFL}{\DIFaddendFL} 
\LetLtxMacro{\DIFOdelbeginFL}{\DIFdelbeginFL} 
\LetLtxMacro{\DIFOdelendFL}{\DIFdelendFL} 
\DeclareRobustCommand{\DIFaddbeginFL}{\DIFOaddbeginFL \let\includegraphics\DIFaddincludegraphics} 
\DeclareRobustCommand{\DIFaddendFL}{\DIFOaddendFL \let\includegraphics\DIFOincludegraphics} 
\DeclareRobustCommand{\DIFdelbeginFL}{\DIFOdelbeginFL \let\includegraphics\DIFdelincludegraphics} 
\DeclareRobustCommand{\DIFdelendFL}{\DIFOaddendFL \let\includegraphics\DIFOincludegraphics} 

\begin{document}

\title{Novel Bounds From The Weak Gravity and Festina Lente Conjectures}

\author{Fayez Abu-Ajamieh}
\email{fayezabuajamieh@gmail.com}
\affiliation{Formerly UC Davis, Davis CA, USA}%

\author{Pratik Chattopadhyay}
\email{pratikpc@gmail.com}
\affiliation{School of Physics, The University of Electronic Science and Technology of China, Chengdu, China}%

\author{Nobuchika Okada}
\email{okadan@ua.edu}
\affiliation{Department of Physics and Astronomy, University of Alabama, Tuscaloosa, USA}%

\author{Roman Pasechnik}
\email{roman.pasechnik@fysik.lu.se}
\affiliation{Department of Physics, Lund University, Lund, Sweden}%

\author{Zhi-Wei Wang}
\email{zhiwei.wang@uestc.edu.cn}
\affiliation{School of Physics, The University of Electronic Science and Technology of China, Chengdu, China}%

\begin{abstract}
We demonstrate that the Weak Gravity Conjecture (WGC) and the Festina-Lente Conjecture (FLC) yield novel bounds on any hypothetical fifth force and milli-Charged Particles (mCPs), as well as on the scale of inflation and on the effective Higgs quartic interaction. In particular, we find that combining the FLC with inflation leads to stronger bounds on mCPs than what the simple application of the FLC provides. Furthermore, we have explored the implications of naturalness on both the FLC and WGC, and have found that these conjectures place a lower limit on the charge of a $U(1)$ gauge group.
\end{abstract}

\maketitle

\section{Introduction}

It is well  known that not all Quantum Field Theories (QFTs) that are consistent at low energies, can have ultraviolet (UV) completions compatible with quantum gravity. QFTs that can be UV completed to include gravity are said to ``live'' in the landscape, whereas those that cannot are considered to belong to the swampland. The main objective of the Swampland program \cite{Vafa:2005ui, Brennan:2017rbf, Palti:2019pca, vanBeest:2021lhn} is to qualify QFTs that can be UV completed to incorporate gravity. This program is supported by a multitude of pieces of evidence from string theory, in addition to other considerations, such as the absence of global symmetries in quantum gravity \cite{Kallosh:1995hi}.

A cornerstone of the Swampland program is the Weak Gravity Conjecture (WGC) \cite{Arkani-Hamed:2006emk}, which roughly states that ``gravity is the weakest force''\footnote{For attempted formulations of the WGC for scalars, see \cite{Palti:2017elp, Lust:2017wrl, Gonzalo:2019gjp, Heidenreich:2019zkl, Benakli:2020pkm, Abu-Ajamieh:2024xic}.}. More precisely, the WGC states that in order for charged black holes in an\textit{asymptotically-flat} spacetime (known as Reissner-Nordstr\"{o}m (RN) black holes) to be able to decay and avoid super-extremality\footnote{For an RN black hole with mass $M$, electric charge $Q$ and magnetic charge $P$, the event horizon is located at $r_{\pm} = GM \pm \sqrt{G^{2}M^{2} - G (Q^{2}+P^{2})}$ ($G$ is the gravitational constant). The black hole is called extremal when $G M^{2} = Q^{2}+P^{2}$, whereas for $G M^{2} < Q^{2}+P^{2}$, it is called super-extremal. For a super-extremal black hole, the metric is regular everywhere except at the (naked) singularity found at $r=0$. This violates the Cosmic Censorship Conjecture.}, the particle with the \textit{smallest mass-to-charge ratio} $m/q$ (with $q$ being the charge under any exact local $U(1)$ symmetry) must obey the following inequality
\begin{equation}
\label{eq:WGC}
m \leq \sqrt{2} g q M_{P}\,,
\end{equation} where $g$ is the gauge coupling and $M_{P}=1/\sqrt{8\pi G} = 2.43 \times 10^{18}$ GeV is the reduced Planck mass. In (quasi-)de Sitter spacetime the relevant solutions are RN-dS black holes, and the WGC in de Sitter has been analyzed explicitly in 
\cite{Antoniadis:2020xso}.\hskip0pt
In the small-H limit, and for black holes whose Schwarzschild radius is much smaller than the Hubble radius $(r_s \ll H^{-1})$, the RN-dS extremality condition reduces to the flat-space form above, which we will use as a local approximation in the quasi-de Sitter regimes considered in this work. Eq.~(\ref{eq:WGC}) is called the electric form of the WGC and should also apply to magnetic monopoles, i.e., $m_{\text{mag}} \leq \sqrt{2} g_{\text{mag}} q_{\text{mag}} M_{P}$. On the other hand, $g_{\text{mag}} = 1/g_{\text{el}}$, and the mass of the magnetic monopole is at least of the order of the magnetic field that it generates, or specifically, $m_{\text{mag}} = \Lambda/g_{\text{el}}^{2}$ on dimensional grounds in terms of a natural cutoff scale $\Lambda$ where the $U(1)$-based effective field theory (EFT) breaks down. Combining the above statements, one deduces the following constraint
\begin{equation}
\label{eq:WGC_magnetic}
\Lambda \lesssim  g q M_{P}\,.
\end{equation}
also known as the magnetic form of the WGC. For instance, considering the hypercharge $U(1)_Y$ symmetry of the Standard Model (SM), Eq.~(\ref{eq:WGC_magnetic}) suggests that the scale of New Physics (NP) is at $\sim 10^{17}$ GeV \cite{Arkani-Hamed:2006emk}. The WGC has several interesting phenomenological implications, see for instance \cite{Shirai:2019tgr, Abu-Ajamieh:2024gaw, Abu-Ajamieh:2024woy,Abu-Ajamieh:2024egb}, and see \mbox{
\cite{Harlow:2022ich}}\hskip0pt
for a comprehensive review of the WGC and its significance.

Another proposed conjecture within the Swampland program is the Festina-Lente Conjecture (FLC)~\cite{Montero:2019ekk, Montero:2021otb}, which states that the mass of \textit{every} particle charged under a given $U(1)$ gauge group must satisfy the following inequality:
\begin{equation}
\label{eq:FL}
    m^{4} \geq 6(q g M_{P}H)^{2} = 2(g q)^{2}V\,,
\end{equation}
where $H$ is the Hubble scale\footnote{The FLC applies only in dS or quasi-dS space. Hence, the Hubble scale in Eq.~(\ref{eq:FL}) could be at any cosmological epoch where the space is either dS or quasi-dS, including the universe during inflation.}, and $V$ is the vacuum energy density. 

The main argument of the FLC relies on the evaporation of charged black holes in dS spacetime. More specifically, extremal charged black holes in dS space, where the Schwarzschild radius and the cosmological radius coincide, are called Nariai black holes \cite{Nariai:1950, Romans:1991nq}. These black holes lose their charge via the Hawking radiation, in addition to the Schwinger pair production. The Schwinger pair production rate is given by\DIFaddbegin \DIFadd{:
}\DIFaddend \begin{equation}
\label{eq:Schwinger}
\Gamma \sim \exp{\Big( -\frac{m^{2}}{qE}\Big)} = \exp{\Big( -\frac{m^{2}}{qg\sqrt{V}}\Big)}\,,
\end{equation}
where the last step arises from the fact that the typical electric field of a Nariai black hole is of order $E \sim g M_{P} H = g \sqrt{V}$. The basic premise of the FLC is that the Nariai black holes should not discharge too quickly, otherwise they would crunch to a naked singularity, thereby violating the Cosmic Censorship Conjecture. Thus, their discharge via the Schwinger pair production must be exponentially suppressed\DIFaddbegin \DIFadd{, }\DIFaddend leading to Eq.~(\ref{eq:FL}). Another way to look at Eq.~(\ref{eq:FL}) is that a magnetic monopole should lie outside its horizon, i.e.~it should not form a black hole, thereby fixing the prefactor in Eq.~(\ref{eq:FL}). We see that while the WGC places an \textit{upper} bound on the mass of a charged particle (specifically, the particle with the smallest $m/q$ ratio in the spectrum), the FLC places a \textit{lower} bound on the mass of a charged particle (in this case, all charged particles in the spectrum).

The FLC is trivially satisfied in QED of the SM, as can be readily verified by applying Eq.~(\ref{eq:FL}) to the electron, the lightest charged particle in the SM. Nevertheless, the FLC carries significant phenomenological implications studied in detail in \cite{Montero:2021otb}. Namely, it was established, for instance, that any $SU(N)$ must either confine or undergo spontaneous symmetry breaking. In addition, it was demonstrated that the FLC forbids the Higgs potential from developing a minimum at $\Phi = 0$ in the broken phase. In \cite{Ban:2022jgm}, the bounds on millicharged particles (mCPs) arising from the mixing between an additional $U(1)$ and the SM $U(1)_{\text{QED}}$ were derived. Furthermore, in \cite{Lee:2021cor}, the FLC was used to investigate a possible UV vacuum structure of the Higgs field beyond the SM (BSM) and to set bounds on the scale of inflation driven by a BSM Higgs potential. 

In this paper, we investigate some further phenomenological implications of the FLC and the WGC. In particular, we consider the relevance of the FLC to the fifth force searches, where we show that both the FLC and WGC can set strong bounds on any fifth force, either gauge $U(1)$ or Yukawa-type. Applied to hypothetical long-range forces, the combined WGC/FLC logic implies a strong parametric constraint; in the extreme IR (mediator masses $(m\sim H))$ this becomes particularly restrictive. We include this limit only to illustrate the scaling with (H), and we do not interpret it as a robust present-day phenomenological exclusion.
We also investigate the implications of the FLC on the Higgs potential instability and set bounds on the effective quartic coupling of the Higgs boson. In addition, we investigate the implications of the FLC on the scale of inflation. Specifically, we show that by combining the FLC with the requirement that the cosmological reheating does not impact the Big Bang Nucleosynthesis (BBN), stringent bounds can be placed on mCPs, which are stronger than those emerging from a simple application of the FLC presented in \cite{Ban:2022jgm}. We also investigate the implications of naturalness on the FLC combined with the WGC, and show that it sets a lower bound on the charge of a $U(1)$ gauge group.

This paper is organized as follows. In Section II, we present the bounds from both conjectures on hypothetical fifth forces. In Section III, we review the magnetic form of the FLC and discuss some of its basic applications to the Higgs instability, inflation, and mCPs. In Section IV, we explore the role of the FLC in light of naturalness, and finally, we conclude in Section V .

\section{The FLC and fifth force searches}\label{sec:Fifth_force}

A fifth force is generally parametrized as a deviation from the Newtonian potential as
\begin{equation}
\label{eq:5th_force1}
U(r) = - \frac{Gm^{2}}{r}(1+\alpha e^{-\frac{r}{\lambda}}) \,,
\end{equation}
where $\alpha$ parametrizes the strength of the new interaction relative to that of gravity, and the force range $\lambda = 1/M$ (not to be confused with the Higgs quartic coupling) in terms of the mass of the force carrier $M$. Thus, it is possible to use both the WGC and the FLC to set upper and lower bounds on any additional $U(1)$ interaction \footnote{We need to point out here that both the FLC and the WGC are based on the discharge and decay of black holes, which were derived for long-range forces (massless U(1) gauge bosons), however, here we are extending the bounds to massive force carriers and not limiting ourselves to the massless regime. So, we are assuming that the decay of BHs via the Hawking radiation, and their discharge via the Schwinger pair production, are not significantly affected by the force carrier being massive instead of massless, as long the mass is not too large.}. However, there is a caveat when applying the WGC: unlike the FLC, which applies to \textit{every} particle charged under some $U(1)$ in the spectrum, the WGC only applies to the particle with the smallest $m/q$ ratio. Since for any $U(1)$ such a particle must exist to allow for charged black holes to decay, we can proceed with using the WGC in the context of constraining the fifth force. 

\begin{figure}[!t] 
\centering
\includegraphics[width=0.6\textwidth]{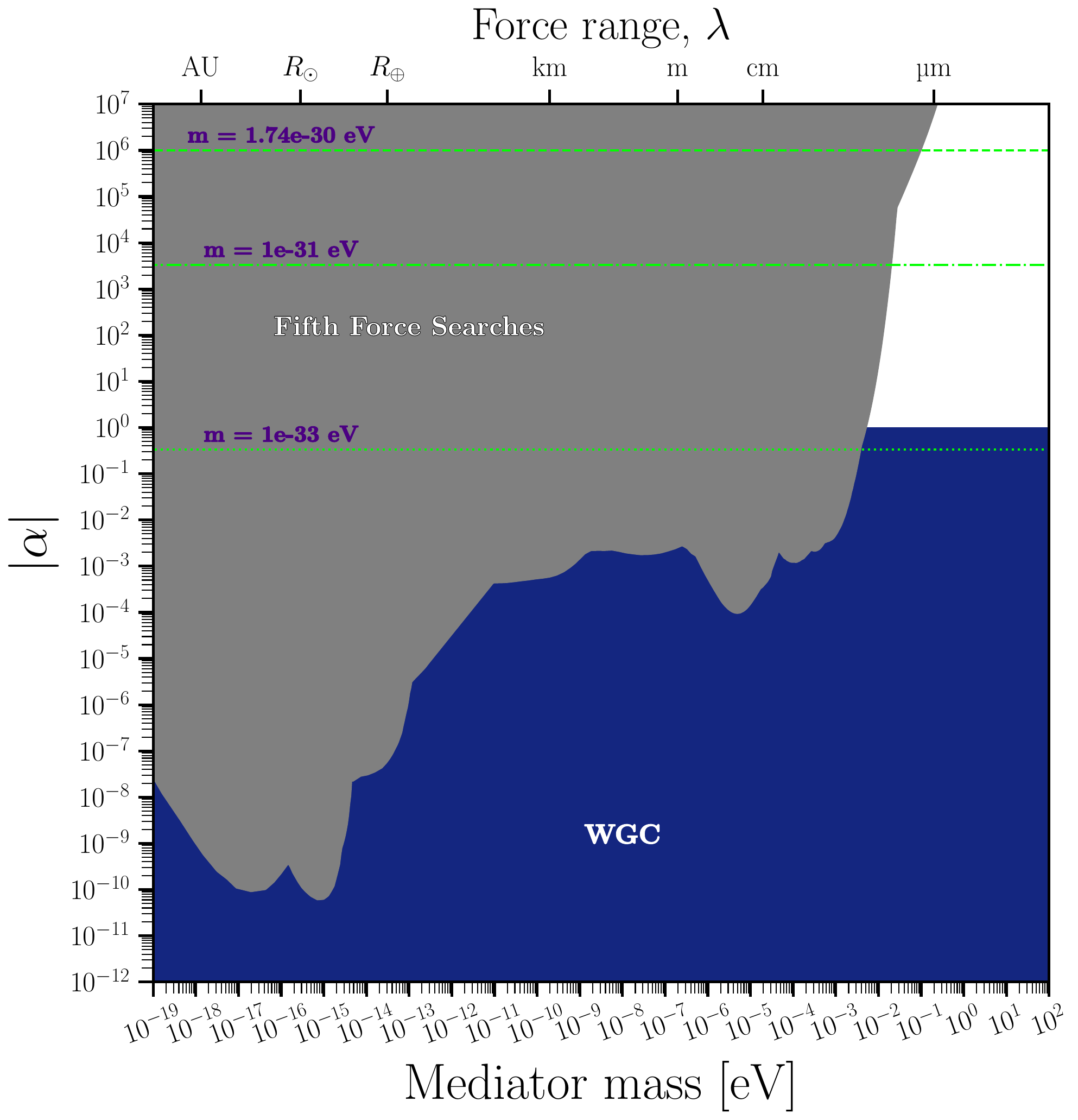}
\DIFdelbeginFL 
{
 Bounds on the fifth force relative strength $\alpha$ provided by the WGC (blue region) and the FLC (green lines) for a broad range of mediator masses $M$ and the corresponding force ranges $\lambda=1/M$. The FLC bounds are shown for a few benchmark particle masses $m$. The region above each line is disfavored for the corresponding benchmark mass. Regions above the curve would be in tension with the FLC for the chosen benchmark. The existing bounds from fifth force searches are shown in gray.} \label{fig1}
\end{figure}

Now, consider a particle of mass $m$, charge $q$, and gauge coupling $g$ under some $U(1)$, whose interaction is mediated by a force carrier of mass $M$ and range $\lambda=1/M$. Then, the $U(1)$ potential can be expressed as
\begin{equation}
\label{eq:5th_force2}
U(r) =-\frac{q^2g^2}{4\pi}\frac{e^{-\frac{r}{\lambda}}}{r} =\frac{-1}{8\pi M_{P}^{2}}\frac{m^{2}}{r}\Big(\frac{2q^{2}g^{2}M_{P}^{2}}{m^{2}}\Big)e^{-\frac{r}{\lambda}}\,.
\end{equation}
Comparing this to Eq.~(\ref{eq:5th_force1}) one finds that $\alpha_{U(1)} = \frac{2q^{2}g^{2}M_{P}^{2}}{m^{2}}$. A comparison with Eq.~(\ref{eq:WGC}) immediately yields the WGC bound, $\alpha_{U(1)} \geq 1$. On the other hand, a straightforward calculation shows that the FLC implies
\begin{equation}
\label{eq:F_5th_force}
\alpha_{U(1)} \leq  \frac{m^{2}M_{P}^{2}}{V} = \frac{m^{2}}{3H^{2}}\,.
\end{equation}
The latter term suggests that the FLC can provide a meaningful bound only for light particles i.e. with masses not much larger than $H \sim 10^{-33}$ eV. Figure~\ref{fig1} shows the bounds implied by the WGC and by the FLC for some benchmark masses on a hypothetical fifth force compared to existing bounds. As the plot shows, both the WGC and the FLC set stringent bounds on the parameter space, and can even disfavor it entirely for masses not much larger than the Hubble scale. Notice that the bounds arising from the FLC are independent of the mass of the force carrier $M=1/\lambda$, as can be seen from Eq.~(\ref{eq:F_5th_force}). We should point out however, that the bounds obtained from the FLC are based on hypothetical masses of the force carriers and only serve as an illustration. In reality, the mass of the force carrier of a potential fifth force would most probably be much larger than $H$, which would lead to a much weaker bound.


\section{The Magnetic FLC and its applications}
\label{sec:magnetic-FLC}

As explained in detail in \cite{Montero:2021otb}, the main argument behind the magnetic form of both the WGC and the FLC is that a magnetic monopole should be a fundamental point-like particle and not form a black hole. More specifically, magnetic monopoles are topologically stable objects that are protected by symmetry\footnote{In more detail, if a gauge group $G$ spontaneously breaks leaving a residual $U(1)$ gauge group $G \rightarrow H \supset U(1)$, then the stability of magnetic monopoles is guaranteed by the residual $U(1)$, and their topological protection arises if the vacuum of the manifold has a nontrivial second homotopy group $\pi_{2}(G/H) \neq 0$.}, which implies they cannot collapse to form a black hole. Thus, the exterior of the monopole ($r \gtrsim 1/\Lambda$, where $\Lambda$ is the energy scale at which the EFT breaks down) should be described by a subextremal magnetically charged Reissner-Nordstr\"om black hole. This implies, according to \cite{Montero:2021otb}, that the following condition must be satisfied
\begin{equation}
\label{eq:polynomial}
    P(r = \Lambda^{-1}, \tilde{M}, \tilde{Q}) = \frac{r^{4}}{\ell_{4}^{2}} + 2\tilde{M}r - n^{2} \tilde{Q}^{2}-r^{2} \leq 0\,,
\end{equation}
where $\tilde{M} = GM$, $\tilde{Q}^{2}n^{2} = \frac{G \pi}{g^{2}}n^{2}$, $\ell_{4}^{2} = 3M_{p}^{2}/V$ , $M$ is the mass of the black hole, $g$ is the coupling of the associated $U(1)$ gauge group, $Q_{n} = 2\pi n$ is the charge of the monopole, $\ell_{4}$ is the 4D de Sitter radius, $V$ is the vacuum energy, and the solution of the polynomial determines the black hole horizon $r_{-}$ and the cosmic horizon $r_{+}$, i.e., $P(r_{\pm}, \tilde{M}, \tilde{Q}) = 0$. In fact, Eq.~(\ref{eq:polynomial}) expresses the condition that $r_{-} \leq \Lambda^{-1} \leq r_{+}$, or alternatively,
\begin{equation}
\label{eq:Mag_FL1}
     \frac{r_{+}}{\ell_{4}} = \sqrt{\frac{1}{6}\Big(1-\sqrt{1-12\tilde{Q}^{2}/\ell_{4}^{2}}\Big)}  \leq  \frac{1}{\Lambda_{4}\ell_{4}}\leq \sqrt{\frac{1}{6}\Big(1+\sqrt{1-12\tilde{Q}^{2}/\ell_{4}^{2}}\Big)}  = \frac{r_{c}^{\text{Nariai}}}{\ell_{4}}\,.
\end{equation}
The latter implies the cutoff $\frac{\tilde{Q}^{2}}{\ell_{4}^{2}} \leq \frac{1}{12}$, which after substituting $\tilde{Q}$ in terms of the coupling $g$ leads to a lower bound,
\begin{equation}
\label{eq:g_bound}
    g^{2} \geq \frac{3}{2}\Big( \frac{H}{M_{P}}\Big)^{2} \,.
\end{equation}
To get a more usable form of the magnetic FLC, we expand Eq.~(\ref{eq:Mag_FL1}) in powers of small $\tilde{Q}^{2}/\ell_{4}^{2}$, which gives to the leading order
\begin{equation}
\label{eq:Mag_FL2}
\Lambda \lesssim \sqrt{8}gM_{p} \leq  <  \frac{2}{\sqrt{3}}\frac{m^{2}}{H} = \frac{2m^{2}M_{P}}{\sqrt{V}}\,,
\end{equation}
where we have used Eq.~(\ref{eq:FL}) in the second relation and $V = 3H^{2}M_{P}^{2}$ in the third. Notice that the first term is essentially the magnetic WGC. Eq.~(\ref{eq:Mag_FL2}) implies that the magnetic FLC can be used to set bounds on the EFT cutoff scale. In what follows, let us consider some useful applications of the magnetic FLC.

\subsection{The Higgs Instability}
\label{sec:4A}

It is well known that the SM Higgs quartic coupling runs into negative values at a scale $\sim 10^{9} - 10^{16}$ GeV, depending on the mass of the top quark. This running is usually interpreted as a sign of NP, and the Higgs potential is assumed to deviate from the SM,
\begin{equation}
\label{eq:BSM_Higgs_V}
V_{\text{eff}}(h) = \frac{1}{4}\lambda(h) h^{4} + \frac{C_{5}}{\Lambda}h^{5} + \frac{C_{6}}{\Lambda^{2}}h^{6} + \cdots \,.
\end{equation}

As we saw above, the magnetic form of the FLC suggests that there is a cutoff scale $\Lambda$ on the EFT associated with any gauged $U(1)$, and it would be interesting to see what this implies for the Higgs vacuum (in)stability. We first define $\lambda_{\text{eff}} \equiv 4V_{\text{eff}}/h^{4}$ as the effective Higgs quartic coupling. Then, the magnetic form of the FLC implies that
\begin{equation}
\label{eq:Mag_FL_Higgs1}
\Lambda \lesssim \frac{2m^{2}M_{p}}{h^{2}\sqrt{\lambda_{\text{eff}}(h)}} \,.
\end{equation}
If we interpret the scale at which the SM quartic supposedly runs into negative values as the scale of NP\footnote{This is a strong assumption, as it implicitly assumes no new physics up to that scale. Therefore, all the bounds derived here are based on this scenario.}, such that $h= \Lambda$, and use the (field-dependent) mass of the electron as being the lightest particle in the SM QED, then we can express Eq.~(\ref{eq:Mag_FL_Higgs1}) as
\begin{equation}
\label{eq:Mag_FL_Higgs2}
\lambda_{\text{eff}}(\Lambda) \lesssim \frac{y_{e}^{4}(\Lambda)M_{p}^{2}}{\Lambda^{2}} \,,
\end{equation}
where we have used $y_{e}(h) = \sqrt{2}m_{e}(h)/h$ and $\lambda_{\text{eff}}(\Lambda)$ is the RGE-evolved coupling of the Higgs quartic at the cutoff scale. As the running of $y_{e}(h)$ is slow in the SM\footnote{Notice that below the scale of NP, any BSM contribution to the $\beta$ function of $y_{e}$ from NP will be suppressed.} and remains $\mathcal{O}(10^{-6})$, Eq.~(\ref{eq:Mag_FL_Higgs2}) can be used to set bounds on $\lambda_{\text{eff}}$ as a function of the scale of NP $\Lambda$.
\begin{figure}[!t] 
\centering
\includegraphics[width=0.8\textwidth]{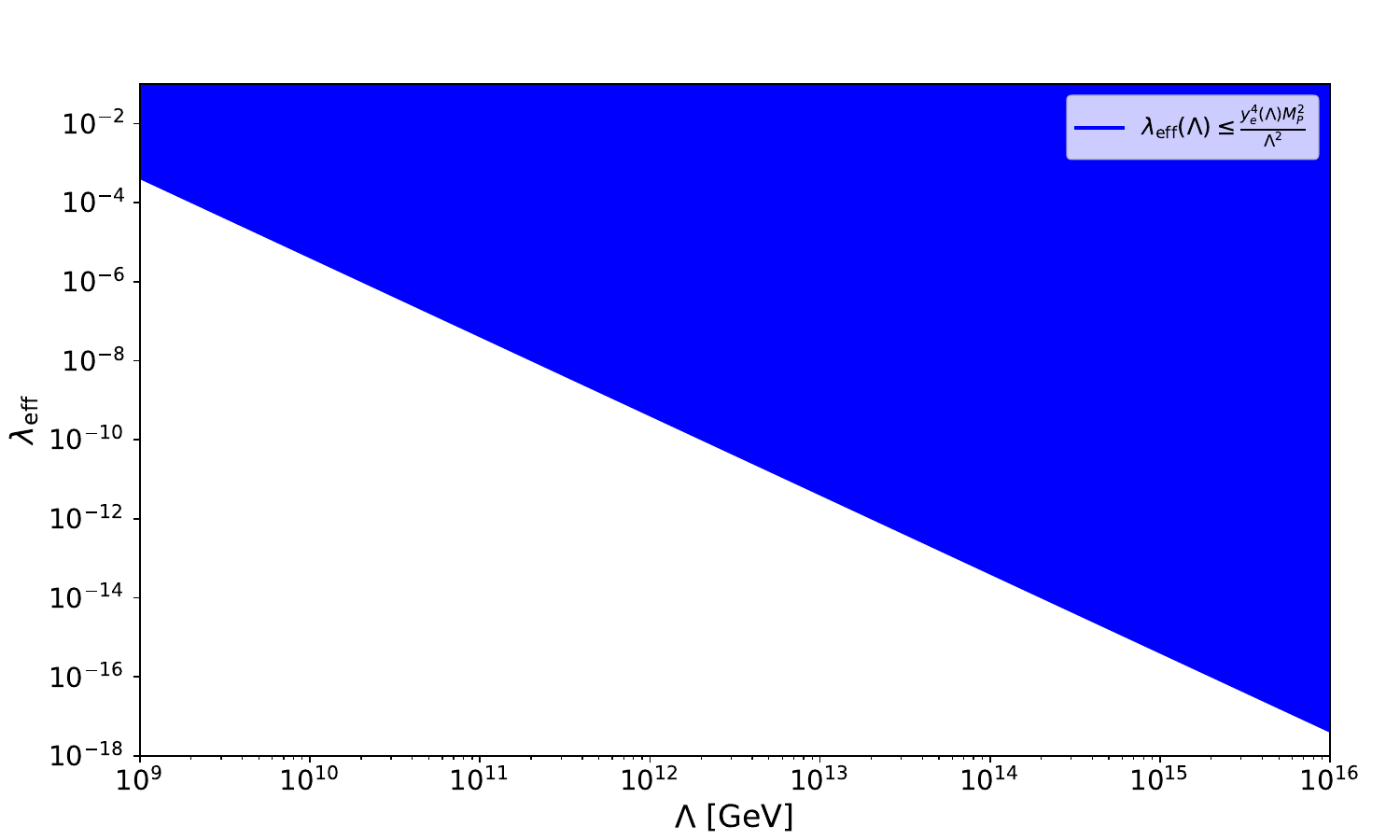}
\caption{Upper bound on $\lambda_{\text{eff}}$ as a function of the scale of NP $\Lambda$.}
\label{fig2}
\end{figure}

This bound is shown in Figure~\ref{fig2}. If we use $\Lambda \sim 10^{9}$ GeV to be conservative, then we obtain $\lambda_{\text{eff}} \lesssim 4.4 \times 10^{-4}$, whereas if we use $\Lambda \sim 10^{16}$ GeV as the upper limit on the scale of NP, then we obtain $\lambda_{\text{eff}} \lesssim 4.4 \times 10^{-18}$. This limit is weaker than $\lambda_{\text{eff}} \lesssim O(10^{-22})$ obtained in \cite{Lee:2021cor}. 

\subsection{Inflation and milli-charged particles}
\label{sec:4B}
Although the electric charge in the SM is quantized, with no experimental evidence so far to indicate any deviation of the charges of SM particles from quantization, it is nonetheless possible for the electric charges to deviate from the SM predictions. More specifically, the requirements of anomaly cancellation and the gauge invariance of the SM Yukawa interactions alone are insufficient to determine charge quantization, and they leave one degree of freedom that allows the electric charge in the SM to deviate from their values by a milli-charge parameter $\epsilon$. In addition, it is not hard to visualize the existence of particles with fractional charges depending on $\epsilon$ that are not quantized. Such particles are known as mCPs.

There are many ways a milli-charge can be introduced, including adding an $SU(2)\times SU(2)$ Dirac fermion with hypercharge $2\epsilon$ \cite{Okun:1983vw}, introducing an additional $U(1)$ gauge group \cite{Holdom:1985ag} that mixes with the SM $U(1)_{\text{em}}$ via kinetic mixing, utilizing the aforementioned remaining degree of freedom in the SM to allow neutrinos to have a milli-charge \cite{Foot:1990uf}, or through introducing non-locality in QED \cite{Capolupo:2022awe, Abu-Ajamieh:2023roj}. Such proposals are interesting for other reasons, such as providing dark matter candidates like the dark photon, which arises in additional $U(1)$ scenarios where the gauge boson is stable or long-lived.

A simple and very straightforward application of the FLC in Eq.~(\ref{eq:FL}) is that it can be used to set constraints on the parameter space of mCPs. This was done in \cite{Ban:2022jgm}, where reasonably stringent bounds on the charge of mCPs  were obtained. Nonetheless, a major limitation on the applicability of the FLC for practical purposes, lies in the smallness of the Hubble scale $H$ at the current epoch of the universe ($H_{0}$). However, this has not always been the case, and $H$ is expected to have been much larger during the epoch of cosmological inflation, $H_{I} \gg H_{0}$. As the FLC is expected to apply  during all epochs of the universe during which the vacuum was dS\footnote{More accurately, and as pointed out in \mbox{
\cite{Venken:2023hfa}}\hskip0pt
, the FLC should only apply in dS space where the vacuum energy is dominant, which implies that $\frac{\rho_{R}}{\rho_{V}} \leq c$, where $\rho_{R}$ is the radiation energy density and $c$ is a constant of $\mathcal{O}(1)$.}, particularly during inflation, it reads ,
\begin{equation}
\label{eq:FL_inflation}
m^{4} \geq 6 (qgM_{P}H_{I})^{2} \,.
\end{equation}

As we will see below, applying the FLC during inflation enables us to obtain much stronger bounds than the naive application of the FLC at the current epoch. To show how this is done, recall that after the end of inflation\DIFaddbegin \DIFadd{, }\DIFaddend the inflaton decays and reheats the Universe.
We define $T_{\rm reh}$ as the temperature at the onset of radiation domination.
Assuming \emph{instantaneous (maximally efficient) reheating}, i.e.\ that the energy density at the end of inflation
is promptly converted into radiation, $\rho_{\rm end}\simeq \rho_R(T_{\rm reh})$, with
$\rho_{\rm end}\simeq 3 M_P^2 H_I^2$ and $\rho_R(T)=\frac{\pi^2}{30}g_{\rm reh}T^4$, we obtain
\begin{equation}
3 M_P^2 H_I^2 \simeq \frac{\pi^2}{30} g_{\rm reh} T_{\rm reh}^4
\quad \Rightarrow \quad
T_{\rm reh}=\left(\frac{90\,M_P^2 H_I^2}{\pi^2 g_{\rm reh}}\right)^{1/4}\,.
\label{inflaton2}
\end{equation}
Here $g_{\rm reh}\equiv g_*(T_{\rm reh})$ is the effective number of relativistic degrees of freedom at reheating.
For a successful Big Bang Nucleosynthesis (BBN), we need $T_{\text{reh}} \gtrsim T_{\text{BBN}}$, which from Eq.~(\ref{inflaton2}) implies a lower bound on the scale of inflation
\begin{equation}
\label{eq:BBN_inf_bound}
H_{I} \gtrsim \frac{\pi \sqrt{g_{\text{reh}}}\,T_{\text{BBN}}^{2}}{3\sqrt{10}M_{P}}\,.
\end{equation}
We can combine this with the FLC during inflation given in Eq.~(\ref{eq:FL_inflation}) to set bounds on mCPs. We obtain (setting $q\equiv \epsilon$)
\begin{equation}
\label{eq:Inf_FL_mCP_bound}
    \epsilon \leq \frac{\sqrt{15}\,m^{2}}{\pi e\sqrt{g_{\text{reh}}}}\frac{1}{T_{\text{BBN}}^{2}} = 1.2 \times 10^{-12} \frac{m^{2}}{\text{eV}^{2}}\,,
\end{equation}
where we have used $T_{\text{BBN}} = 1$ MeV, and $g_{\text{reh}} = 10.75$. 
The bound is shown in Fig.~3, where we also superimpose representative laboratory, astrophysical and cosmological constraints on mCPs for comparison.
As the figure illustrates, the inflationary FLC bound is most relevant in the ultra-light, very small-charge regime, where it provides a complementary constraint, while for larger masses the parameter space is typically dominated by the existing limits shown in the same plot.
For reference, we also display the constraint from the naive present-epoch application of the FLC in Eq.~(3), as discussed in Ref.~[23]; the inflationary bound is parametrically stronger due to $H_I \gg H_0$.
Finally, Eq.~(18) is conservative since it uses the minimal reheating requirement $T_{\rm reh}\gtrsim T_{\rm BBN}\simeq 1~\mathrm{MeV}$; a higher reheating temperature would correspond to a larger $H_I$ and thus further strengthen the bound.

Before we conclude, we emphasize that Eq.~(18) implicitly assumes that the mass and charge of the particle sourcing the Schwinger discharge (e.g.\ the mCP charge carrier) remain approximately constant from the end of inflation through reheating and down to BBN. More generally, if the charge carrier acquires a time-dependent (field-dependent) mass---for instance through a Higgs-like scalar $\phi$ whose background value evolves in time---then the relevant mass during inflation can differ from its late-time value. As a simple example one may have
\begin{equation}
m(\phi)=y\,|\phi| \qquad \text{or} \qquad m^2(\phi)=m_0^2+y^2\phi^2 \, ,
\end{equation}
so that, if $|\phi_{\rm inf}|>|\phi_0|$, one finds $m_{\rm inf}\equiv m(\phi_{\rm inf})>m(\phi_0)$. Since the FLC bound scales as $q_{\max}\propto m^2$, a larger inflationary mass relaxes (weakens) the bound inferred from Eq.(18) when expressed in terms of late-time parameters. Once the field has settled at its minimum $\phi_0$ and the mass becomes time-independent, the bound should be applied using the asymptotic mass $m(\phi_0)$, i.e. Eq.(18) applies again.
\begin{figure}[!t] 
\centering
\includegraphics[width=0.8\textwidth]{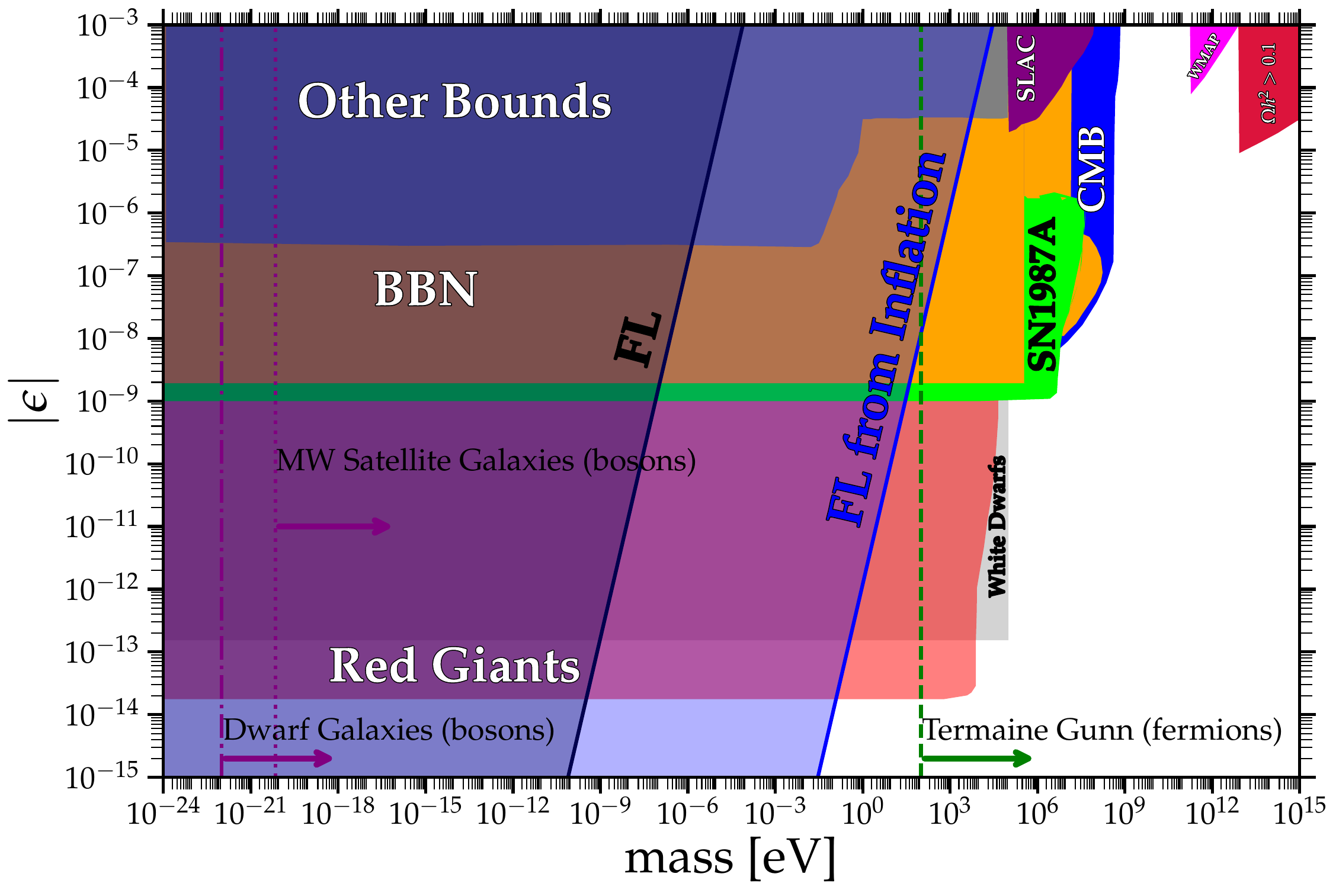}
\caption{Bounds on the mCPs parameter space. The one implied by the FLC from inflation from Eq.~(\ref{eq:Inf_FL_mCP_bound}) is shown in light blue, superimposed with the existing experimental constraints. In addition, we also show the bound implied by  naive application of the FLC as suggested by Eq.~(\ref{eq:FL}) from \cite{Ban:2022jgm}. The inflationary bound is much stronger than the one obtained from the naive FLC application.}
\label{fig3}
\end{figure}

\subsection{The scale of inflation}
\label{sec:4C}

Before we conclude this section, let us take a closer look at Eq.~(\ref{eq:g_bound}), which places a lower bound on the coupling of any gauged $U(1)$. Combining this with the FLC (with $q=1$), we find
\begin{equation}
\label{eq:H_bounds1}
\frac{3}{2}\Big(\frac{H}{M_{P}}\Big)^{2} \leq g^{2} \leq \frac{m^{4}}{6H^{2}M_{P}^{2}} \implies H \leq \frac{1}{\sqrt{3}} m\,.
\end{equation}
For the SM $U(1)_{\text{em}}$, this is trivially satisfied by the electron in the current epoch of the universe. However, it poses a serious problem for inflation once we set $H = H_{I}$, as it would imply an inflation scale being below the mass of the electron, which is not viable. This issue was raised in \cite{Montero:2019ekk, Venken:2023hfa} from the direct application of the FLC, however, unlike those results, Eq.~(\ref{eq:H_bounds1}) is independent of any $U(1)$ coupling and only depends on the mass of the lightest charged particle.

As pointed out in \cite{Montero:2019ekk, Venken:2023hfa}, the tension between the FLC and inflation can be eliminated if the lightest charged particle under the relevant $U(1)$ gauge group is allowed to become heavier. To see this in the case of the SM QED, we substitute $m_{e} = y_{e}v/\sqrt{2}$ in the FLC and combine it with Eq.~(\ref{eq:BBN_inf_bound}). We then obtain
\begin{equation}
 \label{eq:vy bounds}
 y_{e} v \geq \Big( \frac{16\pi^{3}}{15}\alpha_{\text{em}} \,g_{\text{reh}}\Big)^{\frac{1}{4}}T_{\text{reh}}\,,
\end{equation}
and thus we can see that in order to maintain $T_{\text{reh}} \geq T_{\text{BBN}} \sim 1$ MeV, there are only two limiting cases:
\begin{enumerate}
    \item The SM Higgs boson must develop another UV VEV at $v_{\rm UV}$: Setting $g_{\text{reh}}=10.75$ and $y_{e} \sim 10^{-6}$, we find that $v_{\rm UV} \geq 10^{3}$ GeV. This result was obtained in \cite{Lee:2021cor}, however, it was assumed there that $T_{\text{BBN}} \sim 10$ MeV yielding $v_{\rm UV} \geq 10^{4}$ GeV;
    \item The other possibility is for $y_{e}$ to run faster than in the SM, which implies that there must be more degrees of freedom contributing to the $\beta$-function of $y_{e}$. Using $v = 246$ GeV, Eq.~(\ref{eq:vy bounds}) implies that\footnote{Here, we assume an electroweak symmetry broken phase at $\Lambda_{\text{inf}}$.} $y_{e}(\Lambda_{\text{inf}}) \gtrsim 5.16 \times 10^{-6}$.
\end{enumerate}

\section{The FLC and Naturalness}

In this section, we want to investigate the implications of combining naturalness with the FLC. More specifically, we will extend the treatment presented in \cite{Cheung:2014vva} for the WGC , to the FLC. Consider the scalar QED Lagrangian
\begin{equation}
\label{eq:scalar_QED}
    \mathcal{L} = -\frac{1}{4}F_{\mu\nu}F^{\mu\nu} + |D_{\mu}\phi|^{2} - m^{2}|\phi|^{2} -\frac{\lambda}{4}|\phi|^{4},
\end{equation}
where $D_{\mu} = \partial_{\mu} + i q A_{\mu}$. The 1-loop mass correction from the $\phi$ and $A$ loops is given by
\begin{equation}
    \delta m^{2} = \frac{\Lambda^{2}}{16\pi^{2}}(a q^{2} + b \lambda),
\end{equation}
where $\Lambda$ is the cutoff scale. Naturalness implies that $\delta m^{2} \lesssim m^{2}$. Thus, we can use $\delta m^{2}$ in the FLC instead of $m^{2}$. This implies a lower limit on $\Lambda$: 
 \begin{equation}
\label{eq:FL_Lambda_bound}
\Lambda \geq \frac{4\pi\sqrt{\sqrt{6}M_{P}H}}{\sqrt{a q + b \lambda/q}}.
\end{equation}
On the other hand, the WGC places an upper bound on $\Lambda$ as discussed in detail in \cite{Cheung:2014vva}. Combining the two results, we have
\begin{equation}
\label{eq:WGC_FL_Lambda_bound}
\frac{4\pi\sqrt{\sqrt{6}M_{P}H}}{\sqrt{a q + b \lambda/q}} \leq \Lambda \leq \frac{4\pi M_{P}}{\sqrt{a + b \lambda/q^{2}}},
\end{equation}
which implies a lower bound on the $U(1)$ charge $q \geq \sqrt{6}H/M_{P}$. This is identical to the result obtained in \cite{Montero:2021otb} from the combined application of the  FLC  and the WGC. However, here we obtain the same result from the consideration of naturalness. If we use the current value of $H_{0}$, then we get a bound $q \gtrsim 10^{-60}$. However, we can obtain stronger bounds from inflation (see Eq.~(\ref{eq:BBN_inf_bound}))\footnote{We should point out that all such bounds should be understood as lower limits only, as opposed to describing the phenomenology of actual particles.}
\begin{equation}
\label{eq:inf_q_bound}
q \gtrsim \frac{\pi \sqrt{6g_{\text{reh}}}\,T_{\text{BBN}}^{2}}{3\sqrt{10}M_{P}^{2}} \simeq 4.5 \times10^{-43}.
\end{equation}
In addition, we can set an \textit{upper} limit on inflation
\begin{equation}
H_{I} \leq \frac{qM_{P}}{\sqrt{6}} \sim 10^{17} \hspace{1mm} \text{GeV},
\end{equation}
however, this is much weaker than the current experimental bounds of $H_{I} < 10^{13-14}$ GeV obtained from Planck+BICEP/Keck 2018 observations \cite{BICEP:2021xfz}.

\section{Conclusions}
In this paper, we have studied several phenomenological implications of the FLC and the WGC. In particular, we have shown that the FLC and WGC can set stringent bounds on a fifth force. For fifth forces, the WGC/FLC interplay is most restrictive in the extreme-IR limit $(m\sim H)$; we present this limit as a parametric consistency check rather than a present-day exclusion. The phenomenologically relevant bounds we emphasize come from the inflationary application $((H\gg H_0))$ combined with reheating/BBN.
\DIFdelend We have also investigated the implications of the magnetic FLC for the Higgs potential instability, mCPs\DIFaddbegin \DIFadd{, }\DIFaddend and the scale of inflation. In particular, we have found that the FLC sets stringent bounds on the Higgs quartic coupling. Moreover, when the FLC is applied during inflation, the requirement for the reheating temperature to be larger than the BBN temperature can set a stringent bound on the parameter space of mCPs. The latter turns out to be orders of magnitude stronger than what is suggested by the naive application of the FLC. 

We have also shown that the magnetic FLC, as well as the assumption that gravity is the weakest force, can set an upper bound on the scale of inflation that depends only on the mass of the lightest charged particle, which ostensibly leads to tension between the FLC and inflation. However, this tension can be resolved if the lightest charged particle is allowed to become heavier in the UV. We have evaluated the implications of the  interplay between naturalness and the FLC/WGC, and found that this implies a lower limit on the charge of a $U(1)$ gauge group. We point out that the FLC can augment experimental searches in placing limits on the parameter space associated with any $U(1)$ gauge group, whether broken or unbroken.

\section*{Acknowledgment}

R.P.~acknowledges support by the COST Action CA22130 (COMETA).
The work of N.O.~was supported in part by the U.S.~Department of Energy under Grants No.~DE-SC0023713 and DE-SC0012447. Z.W.W. and P.C. are supported in part by the National Natural Science Foundation of China (Grant No. 12475105)

\appendix

\end{document}